\documentclass[12pt, preprint]{aastex}
\usepackage{geometry,amsmath}
\usepackage{natbib}
\shortauthors{Zeng et al.}
\shorttitle{}
\begin{document}

\title{\uppercase{Resolving the Fan-Spine Reconnection Geometry of a Small-Scale Chromospheric Jet Event with the New Solar Telescope}}
\author{Zhicheng Zeng$^{1, 2}$, Bin Chen$^{1, 3}$, Haisheng Ji$^{2, 4}$, Philip R. Goode$^{1, 2}$, and Wenda Cao$^{1, 2}$}
\affil{1.Center for Solar-Terrestrial Research, New Jersey Institute of Technology, 323 Martin Luther King Blvd., Newark, NJ 07102, USA \\
2. Big Bear Solar Observatory, 40386 North Shore Lane, Big Bear City, CA 92314, USA \\
3. Harvard-Smithsonian Center for Astrophysics, 60 Garden Street, Cambridge, MA 02138, USA\\
4. Purple Mountain Observatory, CAS, Nanjing, 210008, China}

\begin{abstract}
Jets present ubiquitously in both quiet and active regions on the Sun. They are widely believed to be driven by magnetic reconnection. A fan-spine structure has been frequently reported in some coronal jets and flares, regarded as a signature of ongoing magnetic reconnection in a topology consisting of a magnetic null connected by a fan-like separatrix surface and a spine. However, for small-scale chromospheric jets, clear evidence of such structures is rather rare, although they are implied in earlier works that show an inverted-Y-shaped feature. Here we report high-resolution (0$''$.16) observations of a small-scale chromospheric jet obtained by the New Solar Telescope (NST) using 10830~\AA{} filtergrams. Bi-directional flows were observed across the separatrix regions in the 10830~\AA{} images, suggesting that the jet was produced due to magnetic reconnection. At the base of the jet, a fan-spine structure was clearly resolved by the NST, including the spine and the fan-like surface, as well as the loops before and after the reconnection. A major part of this fan-spine structure, with the exception of its bright footpoints and part of the base arc, was invisible in the extreme ultraviolet and soft X-ray images (observed by the Atmosphere Imaging Assembly and the X-Ray Telescope, respectively), indicating that the reconnection occurred in the upper chromosphere. Our observations suggest that the evolution of this chromospheric jet is consistent with a two-step reconnection scenario proposed by \citet{Torok2009}.
\end{abstract}


\section{INTRODUCTION}

Jets are a phenomenon of collimated plasma ejecta shooting out from a localized region in the low solar atmosphere. They have been observed in optical, (extreme) ultraviolet (UV/EUV), and X-ray with various sizes, speeds, and durations \citep{Savcheva2007, Nishizuka2011, Tian2014}. Jets usually show a strong tendency to recur, suggesting their association with a prolonged process of magnetic flux emergence and cancellation at their bases \citep{Kurokawa1993, Liu2004}. Observations \citep{Shibata1992, Canfield1996, Liu2009} and MHD simulations \citep{Archontis2005, Nishizuka2008, Pariat2009} support a scenario in which a jet is driven by magnetic reconnection between newly emerging magnetic fluxes and overlying ambient magnetic fields of opposite polarity. This process could result in a configuration for reconnection at a three-dimensional (3D) magnetic null point, connected by a fan-like separatrix surface and a spine along which the reconnected field lines approach or recede from the null \citep[][see, e.g., Fig.~1 of Liu et al. 2011 and Fig.~5d of this paper]{Lau1990, Antiochos1998}. Signatures of such a ``fan-spine'' topology have been reported in various contexts that involve magnetic reconnection, including X-ray jets \citep{Shibata1994, Shimojo1996}, anemone-like active regions \citep{Asai2008}, and ``circular-ribbon'' flares \citep{Masson09, Wang2012, Liu2015}. 

For small-scale chromospheric jets, however, direct observations of fan-spine structures are rather rare, although the existence of a fan-spine structure is strongly implied by the frequently observed inverted-Y-shaped feature of the jet's base \citep{Shibata2007, Nishizuka2008} and in one case, a dome-shaped jet's base outlined by coronal-rain-like flows \citep{Liu2011}.
 
\ion{He}{1} imaging at 10830~\AA{} has proven to be an excellent method for observing plasma with temperatures characteristic of the upper chromosphere and transition region ($>$20,000 K), since the formation of 10830~\AA{} requires the intermediary of collisional excitation to a high energy level and/or a sufficiently high EUV radiation field serving the same purpose. One mechanism is photo-ionization followed by recombination \citep[the ``PR'' mechanism;][]{Zirin75}, in which neutral helium can be ionized by high-energy photons (shortward of 504~\AA{}) and then recombine to the excited $n=2,3$ levels of orthohelium. Another mechanism appeals to the rapid collisional mechanism (CM) in which atoms are excited, before having a chance to be ionized, by electrons with temperatures higher than those expected in ionization equilibrium \citep{Jordan75,Andretta00,Pietarila04}.
 
Here we exploit high-resolution (0$''$.16) 10830~\AA{} imaging utilizing the New Solar Telescope (NST) to resolve the fan-spine structure in a small-scale chromospheric jet event. The instruments and data reduction are discussed in Section 2. Section 3 presents observational results using the NST and other instruments in EUV and X-rays. In Section 4, we consider implications and interpretation of these data.

\section{OBSERVATION AND DATA REDUCTION}

On 2012 July 8, we used the 1.6 meter aperture NST at the Big Bear Solar Observatory \citep[NST/BBSO;][]{Goode12, CaoWD10b} to observe NOAA active region (AR) 11515, which was located close to the west solar limb. NST's off-axis design eliminates any central obscuration, vastly reducing stray light. High spatial resolution images were obtained by using a broad band filter (bandpass: 10~\AA) containing the well-known TiO lines, as well as a narrow band filter (bandpass: 0.5~\AA) placed in the blue wing of the \ion{He}{1} 10830~\AA{} multiplet. Images in H$\alpha$ line center and blue wing ($-$0.8~\AA{}) were also acquired for comparison. 

The 10830~\AA{} Lyot filter was made by the Nanjing Institute for Astronomical and Optical Technology. This narrow-band filter was tuned to $-$0.25~\AA{} relative to the two blended, strongest components of the multiplet (at 10830.3~\AA), making the filtergram capable of imaging fine details of the chromospheric material and meanwhile, capturing the underlying photospheric features. This is also very helpful for co-aligning with other instruments. A high sensitivity HgCdTe CMOS IR focal plane array camera \citep{CaoWD10a} was employed to acquire the 10830~\AA{} data at a cadence of 10~s. With the aid of high order adaptive optics and speckle reconstruction based post-processing method \citep[the KISIP code;][]{Woger07, Cao10c}, images with diffraction limited angular resolution ($\dfrac{\lambda}{D}$) in the three bands were achieved ($\sim$$0\farcs16$ for 10830~\AA{}). 

This event was observed by the Atmosphere Imaging Assembly \citep[AIA;][]{Lemen12} on board the {\it Solar Dynamic Observatory} \citep[{\it SDO};][]{Pesnell12} and the X-Ray Telescope \citep[XRT;][]{Golub2007} on board Hinode \citep{Kosugi2007}, which covered a broad temperature range of the jet material ($\sim$1 MK to $>$ 10 MK). The {\it Reuven Ramaty High Energy Solar Spectroscopic Imager} \citep[{\it RHESSI};][]{LinRP02} and the Gamma-ray Burst Monitor \citep[GBM;][]{Meegan2009} aboard the \textit{Fermi} Gamma-ray Space Telescope also observed this event in hard X-rays (HXRs). RHESSI images are reconstructed using the PIXON algorithm \citep{Hurford2002} based measurements from detectors 3, 5, 6, 7, and 8 with 40-s integration. White-light continuum images from the Helioseismic and Magnetic Imager \citep[HMI;][]{Scherrer2012} on board SDO were used as the intermediary for co-aligning the AIA and NST images. The accuracy of the co-alignment is expected to be better than $0\farcs6$. 

\section{RESULTS}

The repetitive jet event under study occurred around the east edge of the leading sunspot in AR 11515 between 18:19-18:50 UT. Fig.~1 shows images of the jet in 10830~\AA{} (panels a and x), TiO (panel y), H$\alpha$(panel z), EUV (panel b), and soft X-ray (SXR; panel c). The jet was located near the footpoint of a closed coronal loop system as shown in the XRT Al-thick image (Fig.~1c). The second row of Fig. 1 shows a closer view of the jet's base (green box in Fig.~1a), rotated by 110$^{\circ}$ counterclockwise to an upright orientation.
The jet's base ($\sim$10$\arcsec$ wide) appears as a fan-like structure in the 10830~\AA{} filtergram (Fig.~1x), spanning over the penumbral regions of a spot with positive magnetic polarity (Fig.~1y). This structure can also be distinguished in the H$\alpha$ blue wing images, which are, however, badly saturated thus not shown here. Significantly, this structure is totally absent from all AIA EUV images, indicating that its temperature is well below coronal values. The footpoints of the fan-like loops in 10830~\AA{} coincides very well with footpoint brightenings in AIA 171~\AA\ images (yellow contours in Fig.~1x), suggesting strong heating in the footpoint area.

The evolution of the jet event can be generally divided into two major steps separated by $\sim$20 minutes as delineated in the space-time plot of the AIA 304~\AA\ intensity (Fig.~1d; obtained at a slice shown as the dashed line in Fig.~1a). The onset of each step was characterized by a strong footpoint emission in 304~\AA, which correlated well with impulsive peaks in the 1--8~\AA\ soft X-ray (SXR) derivative (from GOES, the Geostationary Operational Environmental Satellite) and the 12--25 keV HXR count rate from Fermi/GBM. This correlation suggests that the jet's evolution was closely associated with electron acceleration and/or strong plasma heating processes \citep{Kane1979,Fletcher2001b}.

Fig.~2 shows the evolution of the jet in the first step observed in 10830~\AA{}, AIA 304~\AA, and H$\alpha$ line center (an animation is available online: stepI.mpg). The jet is visible in 10830~\AA{} as a dark outgoing ejecta, which appears bright in 304~\AA\ and H$\alpha$. With NST's unprecedentedly high resolution, a fan-like structure is clearly seen at the jet's base in the 10830~\AA{} images, consisting of multiple individual loops. The loop system on the left side of the jet is brighter than that on the right, appearing as an unresolved brightening in the 304~\AA\ images (blue arrows in Fig.~2). The evolution of the fan-like loops suggests that magnetic reconnection is likely occurring between the loop systems on the left- and right-side of the jet. To visualize this, we select a curved slice along the fan structure (dashed line in Fig.~2b) and obtain a space-time plot of the 10830~\AA{} intensity (Fig.~4a). At $\sim$18:21:10 UT, two tracks with opposite slopes (directed by red arrows) appear at the slice location of 8$''$ in the space-time plot (white arrow in Fig.~4a). These two tracks correspond to bi-directional plasma outflows ($\sim$30 km~s$^{-1}$) emanating from a common region. In the 10830~\AA{} image, this region is located near the apex of the fan-like structure between the left and right loop systems (marked by an ``x'' symbol in Fig.~2b), presumably the site of magnetic reconnection. The time when the bi-directional outflows occurred coincides very well with the footpoint brightening seen in 304~\AA\ (c.f., Fig.~1d).

The second step starts from 18:39 UT, $\sim$20 minutes after the first step (an animation is available online: stepII.mpg). In contrast to the first step, the 304~\AA\ intensity at the jet's base is dominated by the emission at the right side of the jet (Fig.~3d-f). In 10830~\AA{} images, again, the AIA brightening near the footpoint is resolved as a bundle of small-scale loops. The fan-like surface is clearly visible at the jet's base, composed of alternating black and white fibrils with thickness of $\sim$0.3$''$ (Fig.~3a). A dark spine could be observed near the center of the jet, which becomes more evident in 10830~\AA{}, H$\alpha$, and AIA EUV after 18:41 UT (indicated by white arrows in Fig.~3). This dark spine extends further downward across the fan structure and rooted at the photosphere, representing as the ``inner spine'' in the fan-spine reconnection geometry (c.f., Fig~5d). The foot of the inner spine is co-spatial with a H$\alpha$ brightening (indicated by a blue arrow in Fig.~3g) and a footpoint X-ray source (observed by RHESSI at 6--12 keV, green contours in Fig.~3d). During the late phase of step 2, bright loops form at the right side of the jet spine in 10830~\AA{} (Fig.~3c), whereas a curtain-like surge (5$''$ across) extends to the left side, which appears dark in 10830~\AA{} and bright in EUV 304 and H$\alpha$. Similar to the first step of the jet event, space-time plots obtained across the jet (Fig.~4b, slice 2 in Fig.~3c) show bi-directional flows ($\sim$20 km s$^{-1}$) emanating from a common site near the apex of the fan structure (``x'' symbol in Fig.~3c), which is likely the location of the magnetic null point.

\section{DISCUSSION AND CONCLUSIONS}

We present high-resolution NST observations of a small-scale recurrent jet event using the \ion{He}{1} 10830~\AA{} filter, which is sensitive to upper chromospheric temperatures. The NST observations were complemented by EUV and X-ray data from SDO/AIA, Hinode/XRT, and RHESSI, covering a broad temperature range from $\sim$1 MK to $>$10 MK.

The NST \ion{He}{1} images reveal detailed structures of the recurrent jet event. In particular, the jet at each step consists of an elongated spine along the direction of ejection and a fan-shaped arc at the base, closely matching the fan-spine reconnection geometry for jet production. The fan-shaped arc at the base is only 10$''$ wide, spanning over the penumbral region of a spot of dominating positive magnetic polarity. According to the reconnection picture, the inner spine of the jet (seen in \ion{He}{1} in absorption) that connects to the magnetic null point should be rooted in a newly emerged magnetic flux with negative polarity. It is difficult, though, to confirm this using HMI line-of-sight magnetogram directly, because the negative polarity region was presumably very small and the event occurred near the limb. However, there is multiple evidence supporting the fan-spine reconnection scenario: First, the root of the inner spine coincides with localized H$\alpha$, EUV, and X-ray emissions at the same location (c.f., Fig.~3, indicated by the blue arrow), suggesting intense heating at the root, by either precipitated particles or thermal conduction, which is associated with magnetic reconnection at the null point. Second, bi-directional plasma flows are observed at the onset of each step. They originate from a common region located near the apex of the fan structure, which is probably the site of the magnetic null point where reconnection occurs. This region is located at $\sim$1800 km above the jet footpoints (the projection effect is minimal as the jet event occurred near the limb), probably embedded in the upper chromosphere. Depending on the local Alfv$\acute{e}$n speed, these bi-directional plasma flows, with an observed speed of $\sim$20-30 km s$^{-1}$, could be either direct reconnection outflows or secondary pressure-driven flows expelled from the reconnection site.

We conclude that the observed features in the two major steps of the jet evolution are consistent with the two-step magnetic reconnection scenario of jets proposed by \citet{Torok2009}, as depicted in the two schematic cartoons in Fig.~5: The first step starts when reconnection occurs between the emerged negative magnetic flux and the ambient, unipolar field (Fig.~5c). Reconnected loops shrink downward to the left side of the null point and are heated by the released energy (pink patch). The energy released from reconnection also drives heated plasma from the null point upward along the spine field lines, seen as the jet material in EUV, 10830~\AA{}, and H$\alpha$ (Fig.~2 left column; shaded pink in Fig.~5c). Simultaneously, cool plasma is also ejected from the right side of the jet spine due to magnetic tension force induced by the newly reconnected field lines, also known as the ``slingshot effect'' \citep[][shaded gray in Fig.~5c]{Shibata1994, Canfield1996, Takasao2013}.

The second step corresponds to a fully-developed state of the fan-spine reconnection geometry, when the magnetic loops at the left side of the null point also reconnect with the ambient fields on the right (Fig.~5d). During this step, the reconnected loops are visible on both sides and form a complete fan-shaped structure, separated by an inner spine near the center (Fig.~5b). The dark and bright fibril-like loops consisting of the fan structure (in 10830~\AA{}) represent a mixture of heated and cooled loops formed by the repeated reconnection process at the magnetic null point (Fig.~5b). The reconnection also produces accelerated particles and/or thermal conduction fronts, which propagate along the inner spine and result in strong footpoint X-ray emissions (blue patch in Fig.~5d). 

We attribute our clear detection of the fan-spine structure to not only NST's superb sub-arcsec resolution, but also the special emission conditions for the 10830~\AA{} line. The fan is completely absent in all AIA and XRT bands but clearly visible in \ion{He}{1}, suggesting that the fan should consist of plasma of upper chromospheric temperatures ($\le$20000 K) to which AIA and XRT are not sensitive. The \ion{He}{1} emission is interpreted as being from the radiation of helium atoms in a metastable level excited by thermal collisions (i.e., the CM mechanism). The PR mechanism of \ion{He}{1} emission is not favored for our case since there is insufficient EUV emission in the fan. 

In both steps, the outer spine is observed as a collimated jet consisting of both hot and cool material. This is likely associated with the magnetic reconnection site being embedded in the cool chromosphere: plasma is heated and propelled by the released magnetic energy and meanwhile, cool chromospheric plasma is ejected
due to the sling-shot effect of the reconnected field lines. The inner spine, on the other hand, represents field lines directly connecting the opposite polarity footpoint with the magnetic null point. Its dark appearance in all filters  (10830~\AA{}, H$\alpha$, and EUV) suggests a density depletion at the locus of the inner spine. This is possibly associated with strong heating at the null point combined with efficient thermal conduction along the inner spine field line. This scenario is supported by the existence of a $>$10 MK X-ray source at the footpoint of the inner spine. However we could not completely rule out the possibility of the dark inner spine being associated with a cool filament-like structure that is fortuitously aligned with the jet direction, causing an absorption feature in all bands.

To summarize, the sub-arcsecond resolution of the NST, combined with the unique sensitivity of the 10830~\AA{} line to upper chromospheric plasma of certain temperatures, provide a clear view of a chromospheric jet event with a fan-spine geometry. Although numerous cases of chromospheric jets have been reported with an anemone or inverted Y-shape, our observations, for the first time, reveal nearly every element of the fan-spine structure predicted in the theoretical jet model. The evolution of the jet observed in 10830~\AA{}, H$\alpha$, EUV, and X-ray wavelengths is consistent with a two-step reconnection scenario for jet formation. Our results will motivate further studies on chromospheric jets using high-angular-resolution observations, as well as more detailed modelling approaches.

\acknowledgements {}

The BBSO operation is supported by NJIT, US NSF AGS-1250818, and NASA NNX13AG14G grants. The NST operation is partly supported by the Korea Astronomy and Space Science Institute, Seoul National University, and the strategic priority research program of CAS with Grant No. XDB09000000. We thank Antonia Savcheva and Hui Tian for helpful discussions. We acknowledge the support of the US NSF AGS-0847126, NSFC-11333009, NSFC-11428309, AFOSR (FA 9550-15-1-0322), and NASA under contract SP02H1701R from Lockheed-Martin to SAO and contract NNM07AB07C to SAO.

\newpage

\begin{figure}
\epsscale{1.0} \plotone{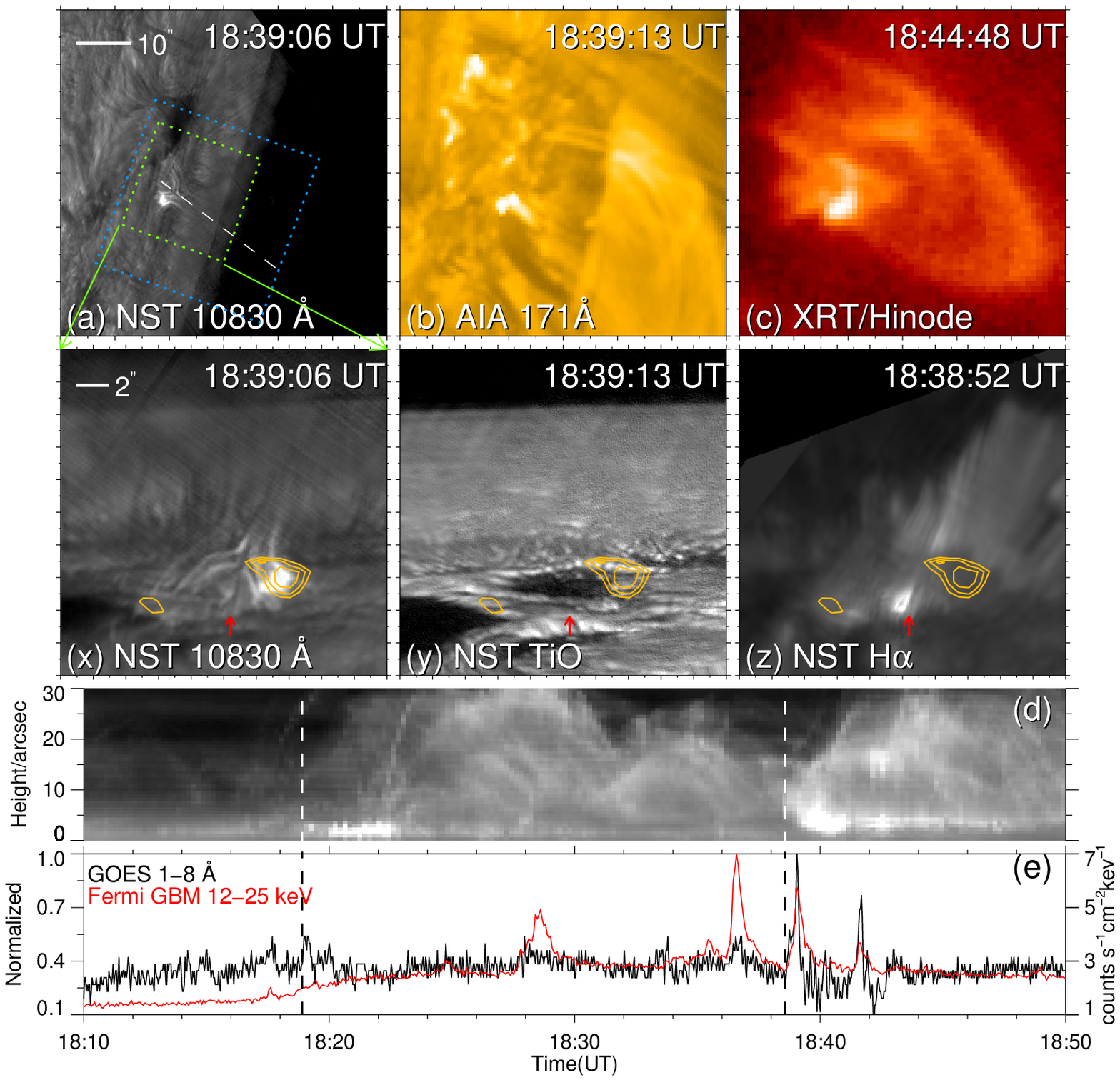}
\caption{The jet event on 2012 July 8: (a) Sample of NST 10830~\AA{} image. The blue box encompasses the field of view (FOV) of the images in Figs.~2-3. (b-c) EUV and X-ray images by AIA/SDO 171 and XRT/Hinode Al-thick, respectively. (a)-(c) have the same FOV of 60$''$$\times$60$''$. (x-z) Closer views of the jet in 10830~\AA{}, TiO, and H$\alpha$ line center, showing the region enclosed in the green box in (a) (rotated counterclockwise by 110$^{\circ}$). The contours show the AIA 171 emission (90\%, 70\%, and 50\% of the peak intensity). Red arrows point to the footpoint of a dark strand in 10830~\AA{} (i.e., the inner spine). (d) Space-time plot made from AIA 304~\AA\ time-series images along a slice in (a) (dashed line). (e) Normalized GOES SXR derivative (black) and Fermi HXR count rate (red). Vertical dashed lines in (d) and (e) denote the start time of each step at 18:18:50 and 18:38:40 UT, respectively.}
\end{figure}

\begin{figure}
\epsscale{1.0} \plotone{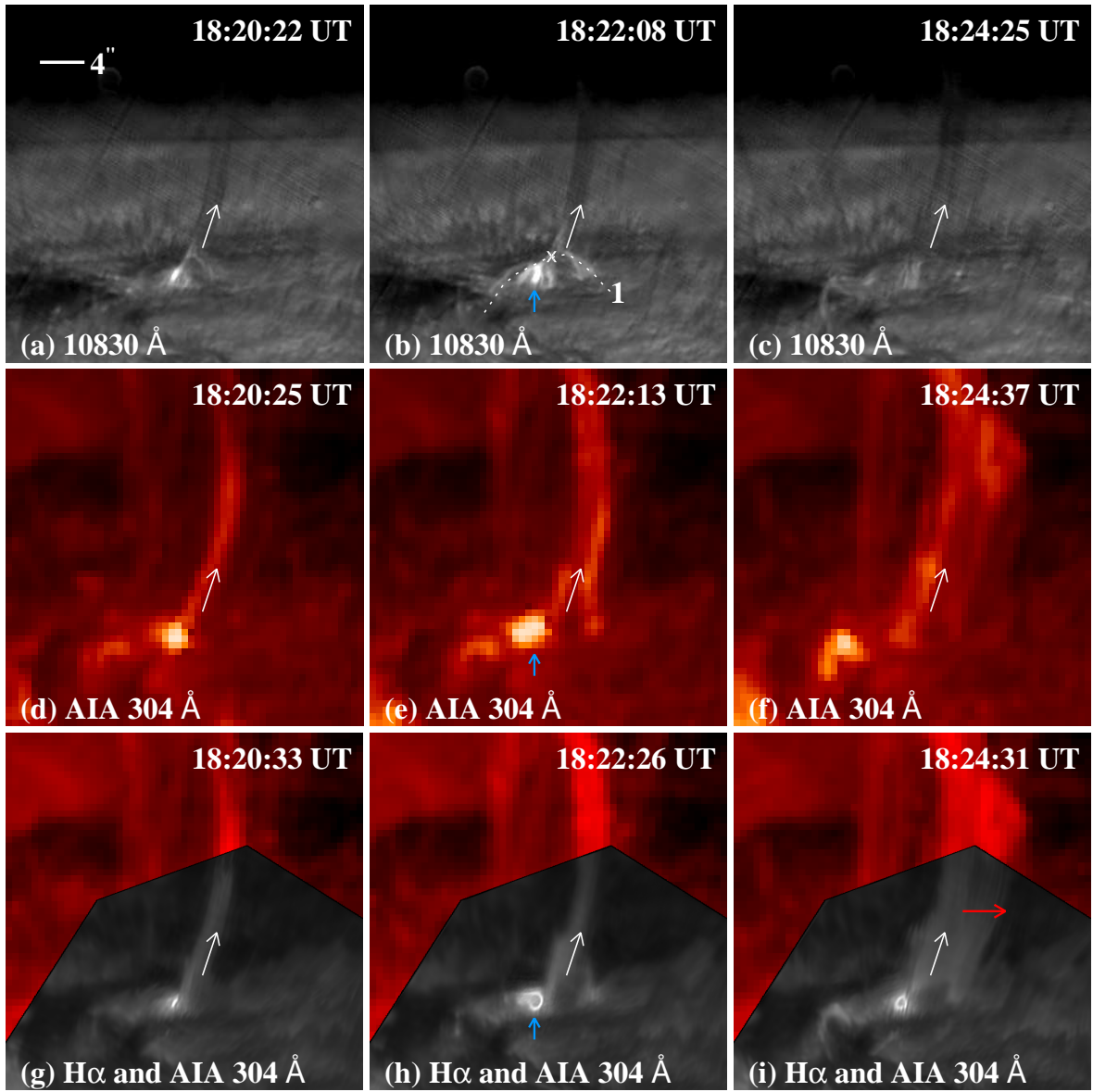}
\caption{Evolution of the jet event during the first step (an animation (stepI.mpg) is available on-line). (a-c) 10830~\AA{} filtergrams. (d-f) AIA 304~\AA\ images. (g-i) Composite of NST H$\alpha$ line center (grey) and AIA 304 images (red). Short blue arrows in the center column depict the ``footpoint'' brightening in AIA 304 (appearing as small-scale loops in 10830~\AA{} at the jet's base). Long white arrows indicate material flowing along the outer spine of the jet. Red arrow in (i) indicates the transverse motion of the H$\alpha$ surge. Slice 1 (dotted line in (b)) is used to obtain the space-time plot in Fig.~4a, with an ``x'' symbol denoting the inferred reconnection site. }
\end{figure}

\begin{figure}
\epsscale{1.0} \plotone{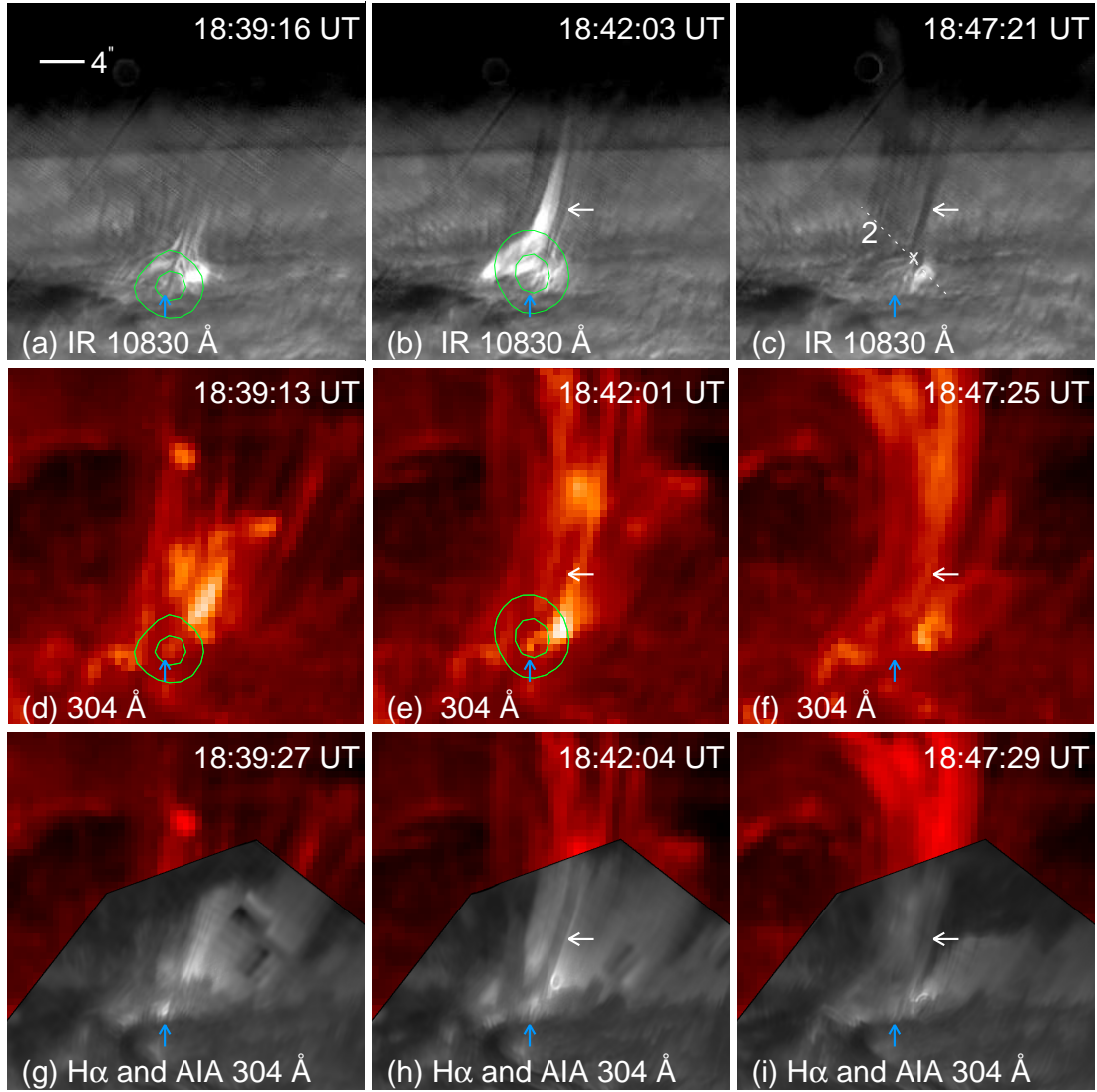}
\caption{Evolution of the jet event in its second stage (an animation (stepII.mpg) is available on-line). (a-c) 10830~\AA{} filtergrams. (d-f)  AIA 304~\AA\ images. Green contours (85\% and 50\% of the peak intensity) in (d) and (e) are RHESSI 6--12 keV X-ray images at 18:39:12 and 18:41:48 UT, respectively. (g-i) Composite of $H\alpha$ line center (gray scale) and AIA 304 images (red). White arrows depict the position of the dark spine while blue arrows point to the footpoint of the inner spine. Slice 2 in (c) is used to obtain space-time plots (Fig.~4b), with an ``x'' symbol denoting the inferred reconnection site.}
\end{figure} 

\begin{figure}
\epsscale{1.0} \plotone{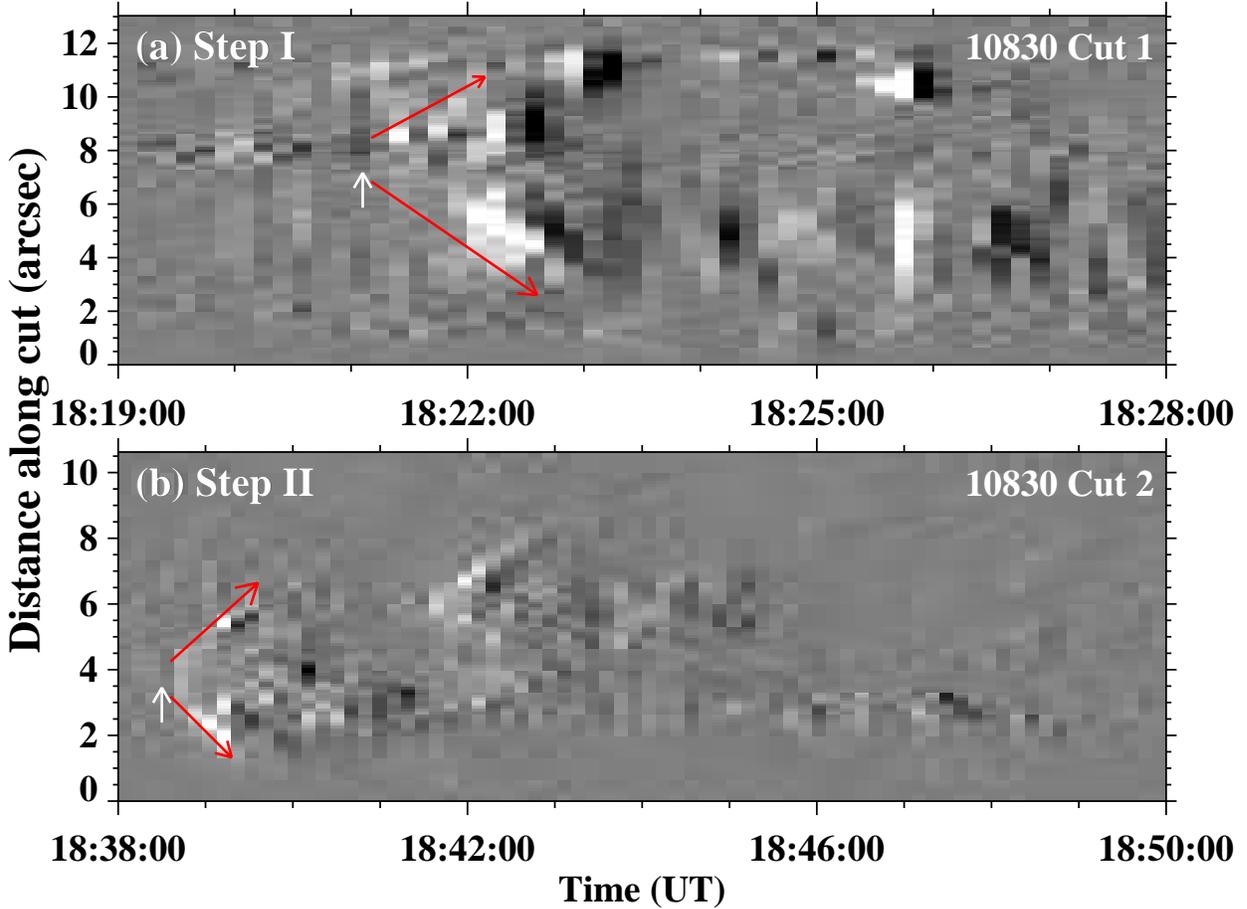}
\caption{Space-time plots showing bi-directional outflows from a common region near the apex of the fan-like structure at the jet's base. (a) and (b) are space-time plots of the 10830~\AA{} intensity obtained along Slice 1 (dashed curve in Fig.~2b) and Slice 2 (dashed line in Fig.~3c) during the first and second step, respectively. Red arrows indicate the bi-directional outflows emerging from a common region (white arrow), which is located near the apex of the fan-like structure (denoted by the ``x'' symbols in Figs.~2b and 3c).
}
\end{figure}

\begin{figure}
\epsscale{1.0} \plotone{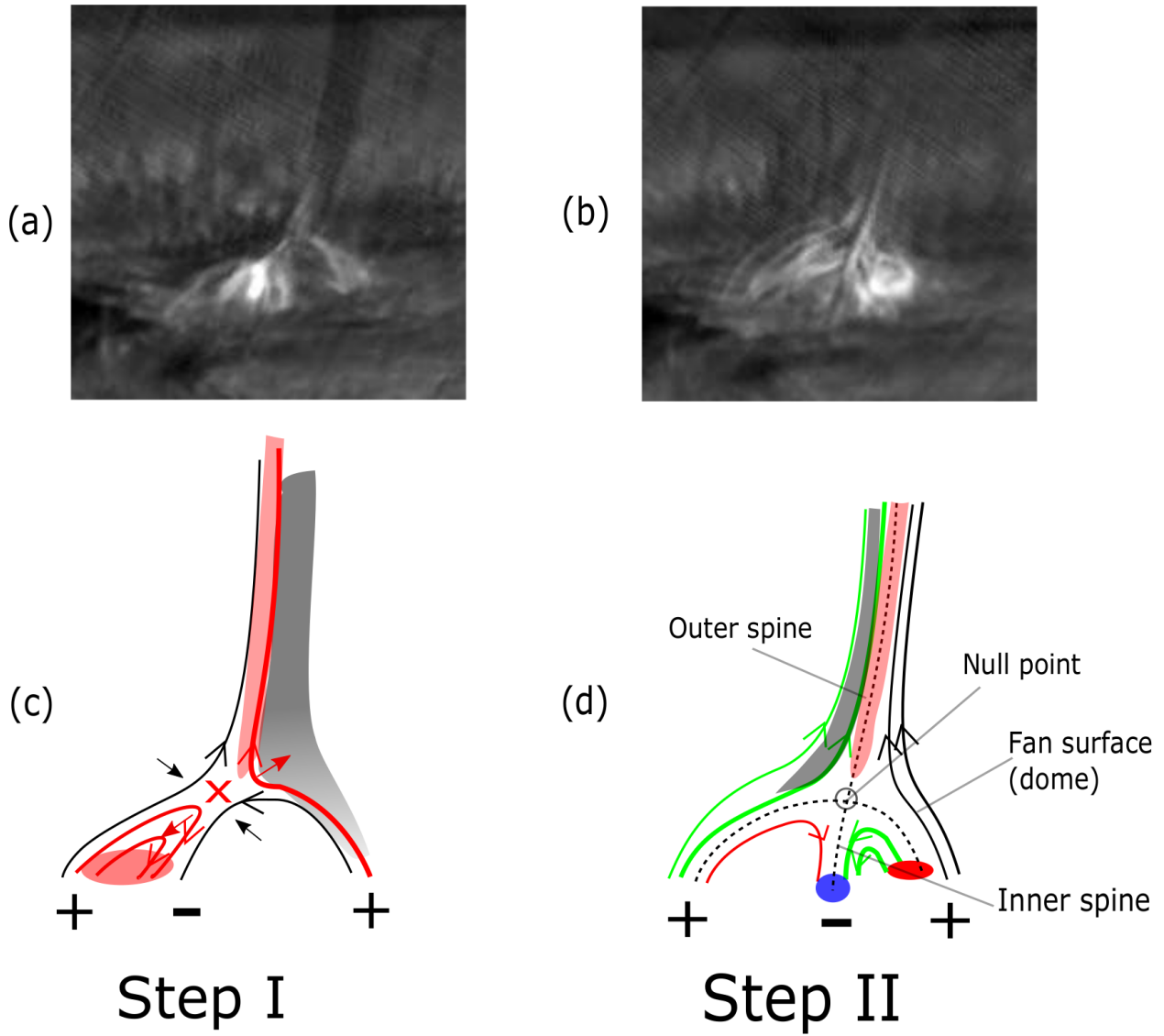}
\caption{(a-b) Representative still images in the first and second step of the jet event. (c-d) Schematic depictions of the event in the two steps. Here red and green lines represent reconnected field lines in the first and second step, respectively. $+$ and $-$ signs denote the magnetic polarity. The pink area near the footpoint represents the footpoint brightening seen in AIA. The jet (in EUV) and the associated surge (in H$\alpha$ and \ion{He}{1}) are shown as the shaded vertical features in pink and grey colors, respectively. The reconnection site is marked by an ``x'' symbol, with small arrows illustrating the reconnection inflows (black) and outflows (red). Blue oval in (d) represents the RHESSI X-ray source at the root of the inner spine. Dotted lines outline the magnetic separatrix layer. }
\end{figure}
\end{document}